\documentclass[aps,prl,twocolumn,showpacs,superscriptaddress,groupedaddress]{revtex4-1}
\usepackage{graphicx}
\usepackage{dcolumn}
\usepackage{bm}
\usepackage{amssymb}
\usepackage{amsmath}
\usepackage{subfigure}

\begin{document}

\title{Optimal Control of Stochastic Magnetization Dynamics by Spin Current}

\author{Yong Wang}
\author{Fu-Chun Zhang}
\affiliation{Department of Physics, The University of Hong Kong, Hong Kong SAR, China}
\begin{abstract}
Fluctuation-induced stochastic magnetization dynamics plays an important role in magnetic recording and writing. Here we propose that the magnetization dynamics can be optimally controlled by the spin current to minimize or maximize the Freidlin-Wentzell action functional of the system hence to increase or decrease the happening probability of the rare event. We apply this method to study thermal-driven magnetization switching problem and to demonstrate the advantage of the optimal control strategy. 
\end{abstract}

\maketitle

One of the key issues in the modern magnetoelectronics is the reliable control of the magnetization dynamics. Compared with the conventional Oersted field generated by the electric current, the spin transfer torque (STT)\cite{STT1,STT2} acting on the magnet by the spin current has been proven to be a more efficient manipulation method as the magnetic structures are miniaturized towards the nanoscale\cite{STTrev1,STTrev2}. Various devices based on this principle has been realized, such as the STT-magnetoresistive random access memory, spin-torque oscillators, spin logic etc\cite{MagRev}. Although the effect of STT on the magnetization dynamics has been extensively studied, its effect on the magnetization fluctuation, which can have a prominent impact on the performance of the spintronics devices\cite{Noise1,Noise2,Noise3,Noise4,Noise5}, has not received enough attentions. Recently, it was demonstrated experimentally that spin current can either suppress or enhance the magnetization fluctuations\cite{Exp}, which coincides with the theoretical studies\cite{LiZhang,WangSham1}.  Furthermore, it raises the interest to control the magnetization fluctuation by spin current and different control strategies have been proposed\cite{Covington,Bando}.     However, there still exists the controversy whether the effective damping should be decreased or increased in order to suppress the magnetization fluctuation\cite{Covington,Bando}, and the choice of control strategies is somewhat arbitrary without solid theoretical foundation\cite{Bando}. Thus deeper understanding the effect of spin current on the stochastic magnetization dynamics and searching for better control strategy are demanded for further applications of the spin current in spintronics devices.  In this letter, we will analyze how the spin current can affect the stochastic magnetization dynamics based on the large deviation principle, and how to optimize the spin current pulses to control the magnetization fluctuation. The application of the idea to the problem of thermal-driven magnetization switching in the presence of STT will be demonstrated as an example. 

The thermal-driven magnetization dynamics under STT is usually given by the stochastic LLG equation\cite{LiZhang} in the case that the quantum effect is negligible\cite{WangSham1}. For given initial magnetization configuration, the fluctuating magnetic field due to the thermal noise will make the trajectories of the magnetzation configuration randomly deviate from its deterministic trajectory when there is no thermal noise. When the noise amplitude is small, most of these random trajectories are around the deterministic trajectory, and only few trajectories can deviate from the deterministic trajectory largely. Thus the magnetization dynamics at long-time scale can be regarded as the combination of the quasi-deterministic motion for most of the time and some large deviation trajectories happened rarely and randomly. Nevertheless, it is because the stochastic magnetization dynamics is mainly governed by these rare trajectories that the control of the stochastic dynamics is reduced to the control of these rare trajectories.   

For simple, we will consider the dynamics of the single-domain magnet with magnetization vector $\mathbf{M}=M_{s}\mathbf{m}$ and volume $V$, where $M_{s}$ denotes the constant magnetization magnitude and $\mathbf{m}$ denotes the unit direction vector. The case for multidomain magnetic structures is more complicated but can still be treated in the similar way.  Generally, the happening probabilities $P[\mathbf{m}(t)]$ of the rare magnetization trajectories $\{\mathbf{m}(t)\}$ in stochastic magnetization dynamics satisfy the large deviation principle\cite{FW,LDPrev,Kohn,Kent}, i.e. taking the exponential form 
\begin{equation}
P[\mathbf{m}(t)]\asymp e^{-S[\mathbf{m}(t),I^{s}(t)]/\epsilon },\label{Prob1}
\end{equation}
where the Freidlin-Wentzell (FW) action functional $S[\mathbf{m}(t),I^{s}(t)]$ depends on the applied spin current $I^{s}(t)$, and the parameter $\epsilon$ reflects the noise amplitude. In order to increase (or decrease) the happening probability of certain magnetization trajectory $\{\mathbf{m}(t)\}$, one can adjust the applied spin current $I^{s}(t)$ to decrease (or increase) the FW action functional. Specially, the maximal (or minimal) happening probability for $\{\mathbf{m(t)}\}$ is achieved by minimizing (or maximizing) the FW action functional. Usually, certain constrait conditions for the spin current should be set up, such as the total consumed energy during the control should be constant, etc. Then the \emph{optimal} spin current $I_{opt}^{s}(t)$ to control the happening probability of the given magnetization trajectory $\{\mathbf{m}(t)\}$ is given by the variational problem\cite{OpCon}
\begin{eqnarray}
\delta\{S[\mathbf{m}(t),I^{s}(t)]+\lambda F[I^{s}(t)]\}=0,\label{var1}
\end{eqnarray}
where we have set the constraint conditions $F[I^{s}(t)]=0$ for the spin current, and $\lambda$ is the Lagrange multiplier.  

The dynamics of the single-domain magnet are driven by the deterministic effective magnetic field given by the micromagnetic energy density, the STT due to the spin current, and the stochastic magnetic field arising from the thermal fluctuation. Explicitly, the effective magnetic field is $\mathbf{H}_{eff}=-\nabla_{\mathbf{m}}E(\mathbf{m})/M_{s}$, with the micromagnetic energy density $E(\mathbf{m})$. The spin current $I^{s}$ is defined by its amplitude $a_{J}$ and spin-polarization vector $\mathbf{P}$, and we will only consider the adiabatic STT\cite{STT1,STT2} although the non-adiabatic STT can be important in some cases\cite{STT3}. The fluctuating magnetic field is assume as $\sqrt{\epsilon}\dot{\mathbf{W}}$, where the amplitude $\epsilon=\frac{2\alpha K_{B}T}{\gamma M_{s}V}$ is proportional to temperature $T$ and Gilbert damping coefficient $\alpha$, and $\dot{\mathbf{W}}$ is the Gaussian white noise process satisfying $\langle\dot{\mathbf{W}}(t)\rangle=0$ and $\langle\dot{W_{i}}(t)\dot{W_{j}}(t')\rangle=\delta_{ij}\delta(t-t')$. Here we have introduced the Boltzmann constant $k_{B}$ and the gyromagnetic ratio $\gamma$. Then the stochastic LLG equation can be written in the compact form\cite{Kent}
\begin{eqnarray}
\dot{\mathbf{m}}=\mathbf{b}(\mathbf{m})+\sqrt{\epsilon}\sigma(\mathbf{m})\dot{\mathbf{W}},\label{LLG}
\end{eqnarray} 
where the vector $\mathbf{b}$ and the matrix $\sigma$ are defined as
\begin{eqnarray*}
\mathbf{b}(\mathbf{m})&=&\sigma\mathbf{H}_{eff}-a_{J}\gamma'K_{S}\mathbf{P},\\
\sigma(\mathbf{m})&=&\gamma^{\prime }(K_{A}+\alpha K_{S}).
\end{eqnarray*}
Here we have $\gamma'=\frac{\gamma}{1+\alpha^{2}}$, and the elements of the antisymmetric matrix $K_{A}$ and symmetric matrix $K_{S}$ are\cite{Kent}
$(K_{A})_{\mu\nu}=\epsilon_{\mu\nu\rho}m_{\rho},\quad(K_{S})_{\mu\nu}=\delta_{\mu\nu}-m_{\mu}m_{\nu}.$  In this case, the expression of the FW action functional $S[\mathbf{m}(t),I^{s}(t)]$ will be\cite{Kohn,Kent} 
\begin{equation}
S[\mathbf{m}(t),I^{s}(t)]=\frac{1}{2}\int_{t_{i}}^{t_{f}}|\sigma^{-1}(\dot{\mathbf{m}}-\mathbf{b})|^{2}dt,\label{action1}
\end{equation}
where $t_{i}$ and $t_{f}$ are the initial and final time for the trajectory $\{\mathbf{m}\}$. Equations (\ref{var1}) and (\ref{action1}) are the theoretical foundations for us to optimally control the targeted trajectory $\{\mathbf{m}\}$ of the sinlge-domain magnet with spin current $I^{s}$. 

For the above optimal control problem, the usual situation is that many targeted trajectories exist. For example, the magnet may switch from one stable configuration to another via different paths. Among these targeted trajectories, the one which minimizes the FW action functional $S[\mathbf{m}(t),I^{s}]$ for fixed spin current $I^{s}$ would be the most probable trajectory, which is the so-called \emph{optimal path}\cite{OpCon}. The optimal path provides the best prediction for the random and rare trajectories happened in practice, and the corresponding happening probability is  $P\sim e^{-V[\mathbf{m}_{opt},I^{s}]/\epsilon}$, where $V[\mathbf{m}_{opt},I^{s}]=\text{min}\{S[\mathbf{m}(t),I^{s}]\}$ is the quasi-potential\cite{Kohn,Kent}.  Thus, the optimal control problem is further reduced to find the optimal spin current $I^{s}_{opt}$ to minimize or maximize the quasi-potential in order to increase or decrease the happening probability of the optimal path.

In principle, the double optimization problem formulated above can be solved by various optimization algorithms. One special but important case is that the spin current $I^{s}$ is weak and can be treated perturbatively in the magnetization dynamics. Then the optimal path $\{\mathbf{m}_{opt}\}$ can be assumed to be independent on the spin current, and the quasi-potential $V[\mathbf{m}_{opt},I^{s}]$ can be approximated to the first order of $a_{J}$ as $V[\mathbf{m}_{opt},I^{s}]=V_{0}[\mathbf{m}_{opt}]+\Delta V[\mathbf{m}_{opt},I^{s}]$, where
\begin{eqnarray}
V_{0}&=&\frac{1}{2}\int_{t_{i}}^{t_{f}}|\sigma^{-1}(\dot{\mathbf{m}}_{opt}-\mathbf{b}_{0})|^{2}dt,\label{action0}\\
\Delta V&=&\int_{t_{i}}^{t_{f}}a_{J}\gamma'\langle \sigma^{-1}(\dot{\mathbf{m}}_{opt}-\mathbf{b}_{0}), \sigma^{-1}K_{S}\mathbf{P}\rangle dt.\label{Daction}
\end{eqnarray}
For convenience, we have introduced the vector $\mathbf{b}_{0}\equiv\sigma\mathbf{H}_{eff}$, and notice that $\sigma$, $K_{S}$, $\mathbf{H}_{eff}$ are all functions of $\mathbf{m}_{opt}$. Thus, the presence of a weak spin current $I^{s}$ gives the change of the initial quasi-potential $V_{0}$ in the amount of $\Delta V$, and the happening probability of the optimal path $\{\mathbf{m}_{opt}(t)\}$ has been exponentially changed by the factor $e^{-\Delta V/\epsilon}$. The optimal spin current $I^{s}_{opt}$ is obtained by minimizing or maximizing $\Delta V$, i.e.
\begin{eqnarray}
\delta\{\Delta V[\mathbf{m}_{opt}(t), I^{s}(t)]+\lambda F[I^{s}(t)]\}=0.\label{var2}
\end{eqnarray}

Thermal-driven magnetization switching is a typical example of the stochastic dynamics described above, and the concept of ``effective temperature" has been introduced to the N\'eel-Brown law to take account into the effect of spin current on the magnetization switching probability\cite{Noise2,Noise3,Noise4,LiZhang}. This fact can be easily verified from Eq.~(\ref{action0}) and (\ref{Daction}) for a weak spin current with constant amplitude, where $\Delta V$ is now proportional to $a_{J}$ and the magnetization switching probability is formally written as $P\sim e^{-(1-\frac{a_{J}}{a_{c}})V_{0}/\epsilon}$.
Here, $a_{c}$ is the critical value for the spin current to switch the magnet, and is given by 
\begin{eqnarray}
a_{c}=\alpha\frac{\int_{t_{i}}^{t_{f}}|\sigma^{-1}K_{S}\mathbf{H}_{eff}|^{2}dt}{\int_{t_{i}}^{t_{f}}\langle\sigma^{-1}K_{S}\mathbf{H}_{eff}, \sigma^{-1}K_{S}\mathbf{P}\rangle dt}.\label{ac}
\end{eqnarray}
To get Eq.~(\ref{ac}), we have used the fact that in the absence of STT, the optimal path $\{\mathbf{m}_{opt}\}$ satisfies the equation\cite{Kohn}
\begin{eqnarray} 
\dot{\mathbf{m}}_{opt}=\gamma'(K_{A}-\alpha K_{S})\mathbf{H}_{eff}.\label{OptPath}
\end{eqnarray} 
From Eq.~(\ref{ac}), it is seen that the value of $a_{c}$ is dependent on the choice of the spin polarization vector $\mathbf{P}$ of the spin current. Especially, if $\mathbf{P}$ is always in anti-parallel with the effective magnetic field $\mathbf{H}_{eff}$, then $a_{c}$ will be negative and the magnetization switching probability $P$ will be decreased exponentially. From Eq.~(\ref{LLG}), one might regard that the Gilbert damping should be effectively enhanced by the spin current in order to stabilize the magnet\cite{Bando}. One the other hand, $\mathbf{P}$ should be parallel with $\mathbf{H}_{eff}$ in order to increase $P$. Furthermore, Eq.~(\ref{ac}) gives a rough estimation of the critical spin current magnitude as $a_{c}\sim\alpha\,\text{max}\{|\mathbf{H}_{eff}|\}$, which is proportional to the Gilbert damping coefficient and the maximal effective field\cite{LiZhang}. 

If the spin current amplitude is assumed to be time-dependent, the optimal pulse shape $a^{opt}_{J}(t)$ to control the optimal path $\{\mathbf{m}_{opt}\}$ can be obtained from Eq.~(\ref{var2}). One natural constraint condition would be that the total dissipation energy of the spin current pulse should be constant, i.e. $\int_{t_{i}}^{t_{f}}a_{J}^{2}(t)dt=\mathcal{E}$. Then the general solution $a_{J}^{opt}(t)$ of Eq.~(\ref{var2}) is given by
\begin{eqnarray}
a_{J}^{opt}(t)-\frac{1}{\lambda}\alpha\gamma'^{2}\langle\sigma^{-1}K_{S}\mathbf{H}_{eff},\sigma^{-1}K_{S}\mathbf{P}_{opt}\rangle=0.\quad\label{ajopt}
\end{eqnarray}
and the Lagrange multiplier $\lambda$ is determined from the constraint condition as
\begin{eqnarray}
\lambda=\pm\alpha\gamma'^{2}[\frac{1}{\mathcal{E}}\int_{t_{i}}^{t_{f}}\langle\sigma^{-1}K_{S}\mathbf{H}_{eff},\sigma^{-1}K_{S}\mathbf{P}_{opt}\rangle^{2}dt]^{\frac{1}{2}}.\quad\quad\label{Lmul}
\end{eqnarray}
This spin current pulse gives the change of the quasi-potential as $\Delta V=-2\lambda\mathcal{E}$. According to Eq.~(\ref{ajopt}), the sign of $\lambda$ is postivie (or negative) when the optimal spin polarization vector $\mathbf{P}_{opt}$ is parallel (or antiparallel) with the effective magnetic field $\mathbf{H}_{eff}$, which will enhance (or suppress) the happening probability of the optimal path $\mathbf{m}_{opt}(t)$ by the factor $e^{2\lambda\mathcal{E}/\epsilon}$.

Based on the theoritical analysis above, we use the stochastic LLG equation to simulate the optimal control of the thermal-driven magnetization switching by spin current numerically. Considering a ferromagnetism film with easy axis along $z$-axis, and demagnetization field direction along $x$-axis, then the energy density $E(\mathbf{m})$ is given as\cite{LiZhang,Bando,Kent}
\begin{eqnarray}
E(\mathbf{m})=-\frac{1}{2}H_{a}M_{s}m_{z}^{2}+2\pi M_{s}^{2}m_{x}^{2}.\label{Eden}
\end{eqnarray}
Here, we assume that no external magnetic field is applied, the anisotropic field is $H_{a}=0.05$~T, and the demagnetization field is $4\pi M_{s}=1.2$~T. Besides, we set the Gilbert damping coefficient $\alpha=0.03$, the temperature $T=300$~K, and the magnet volume $V=1500$~nm$^{3}$. The initial direction of the magnet is assumed as $\mathbf{m}=(0,0,1)$, and the stochastic LLG equation (\ref{LLG}) is simulated with the 4th-order Runge-Kutta method, where the time step is set as 1~ps. We also got the amplitude of the critical spin current numerically as $a_{c}=0.024$ at zero temperature. Notice that we have chosen a relatively small magnet otherwise the switching time will be too long to be numerically simulated.

\begin{figure}[ht]
\includegraphics[scale=0.175]{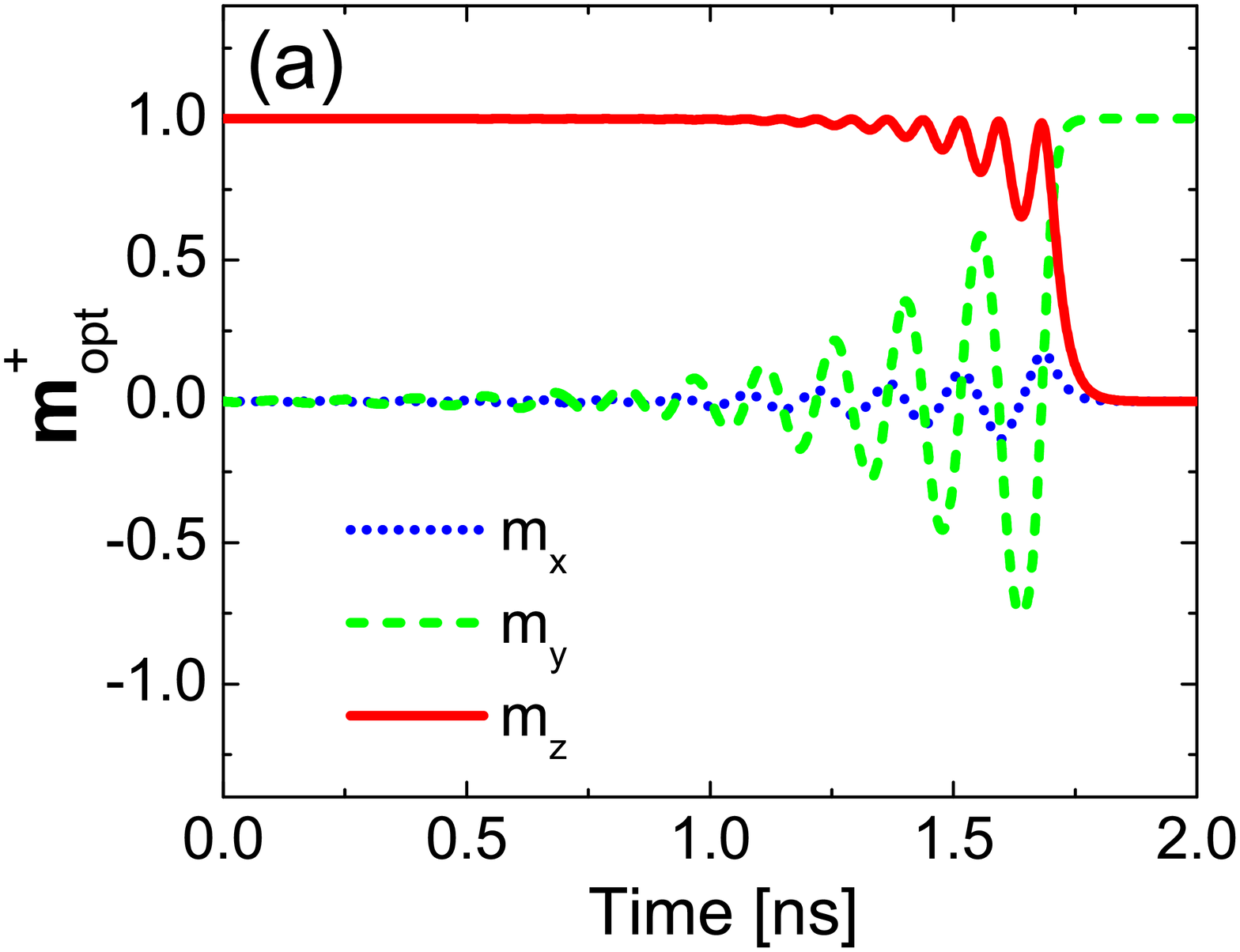}
\includegraphics[scale=0.175]{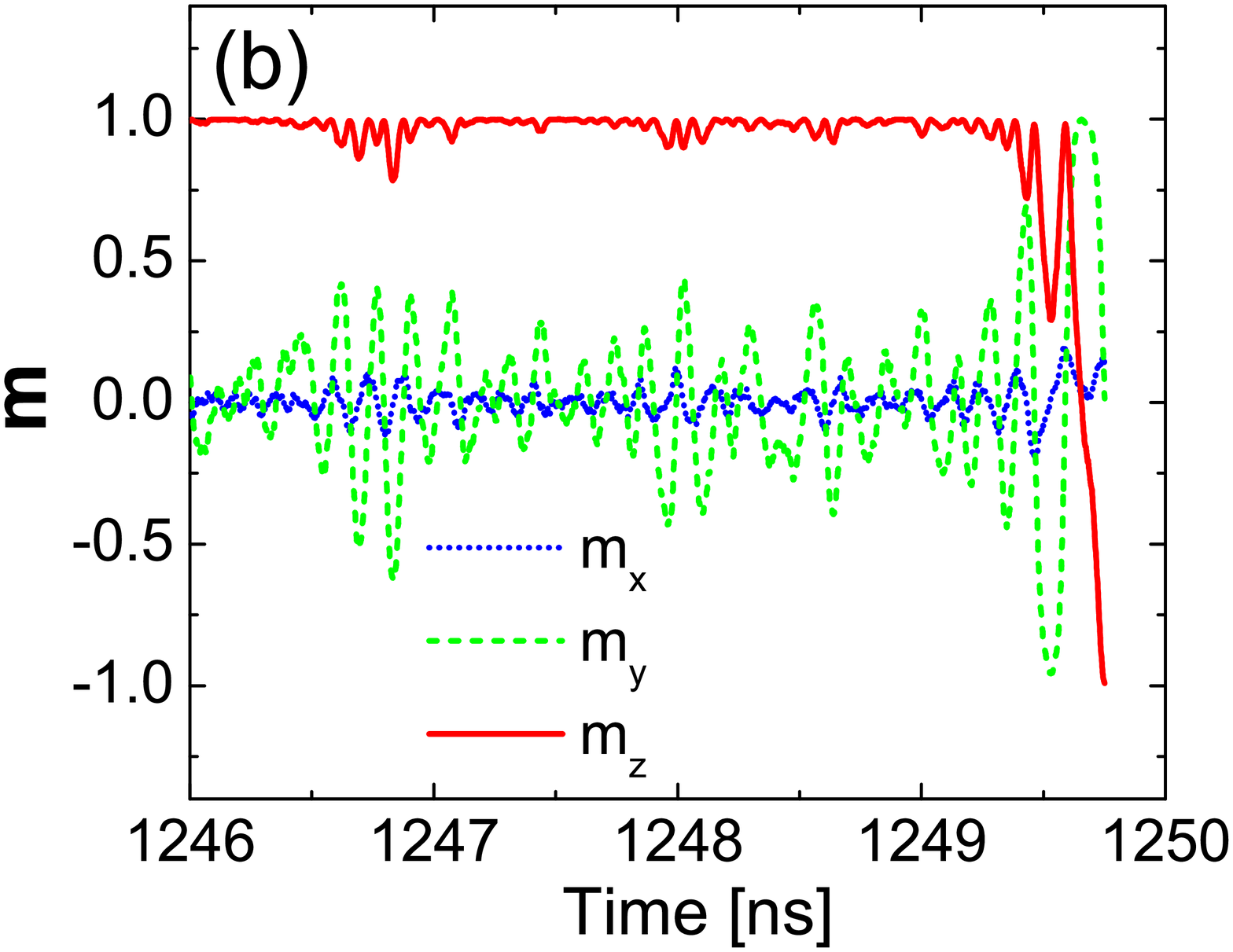}
\includegraphics[scale=0.175]{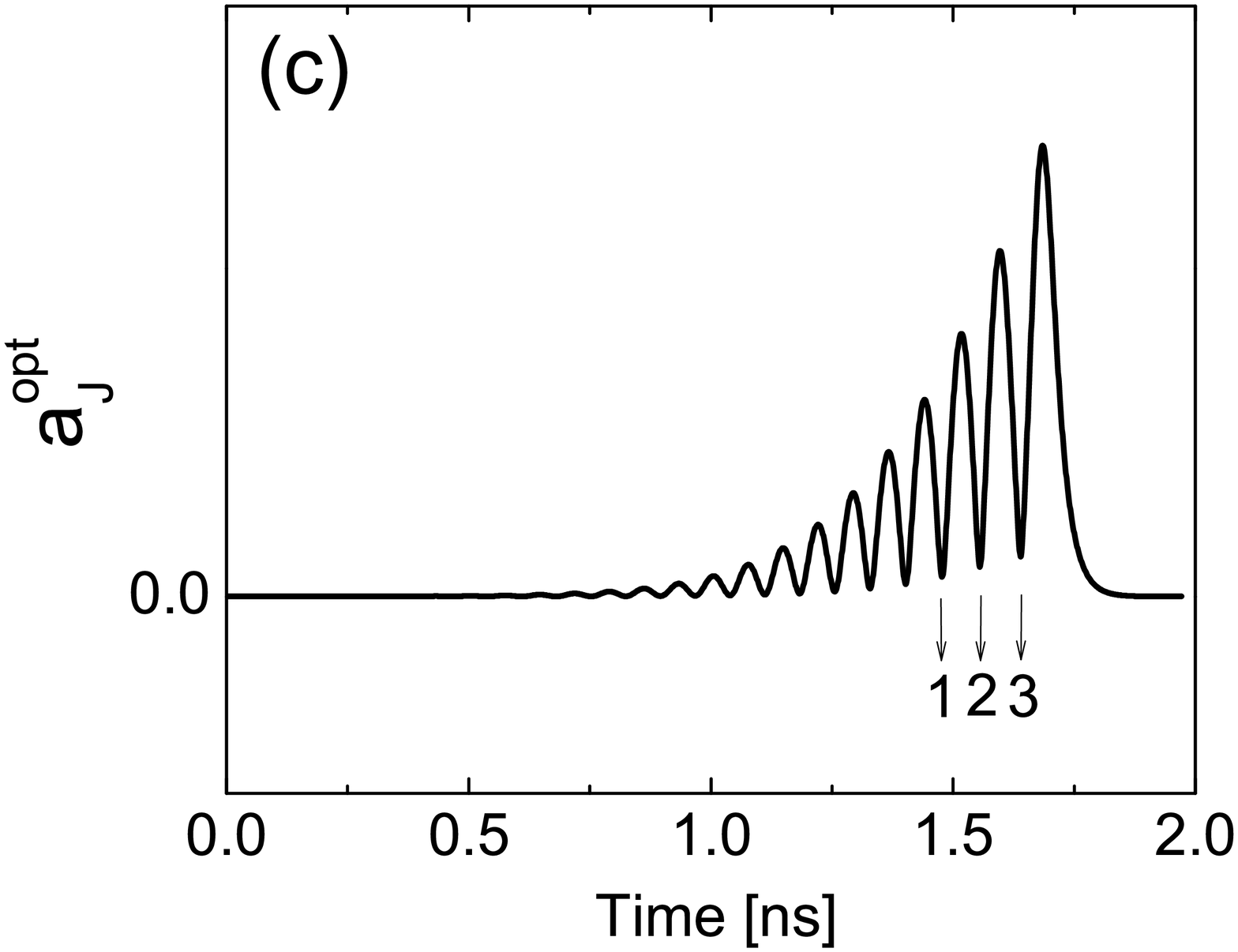}
\includegraphics[scale=0.175]{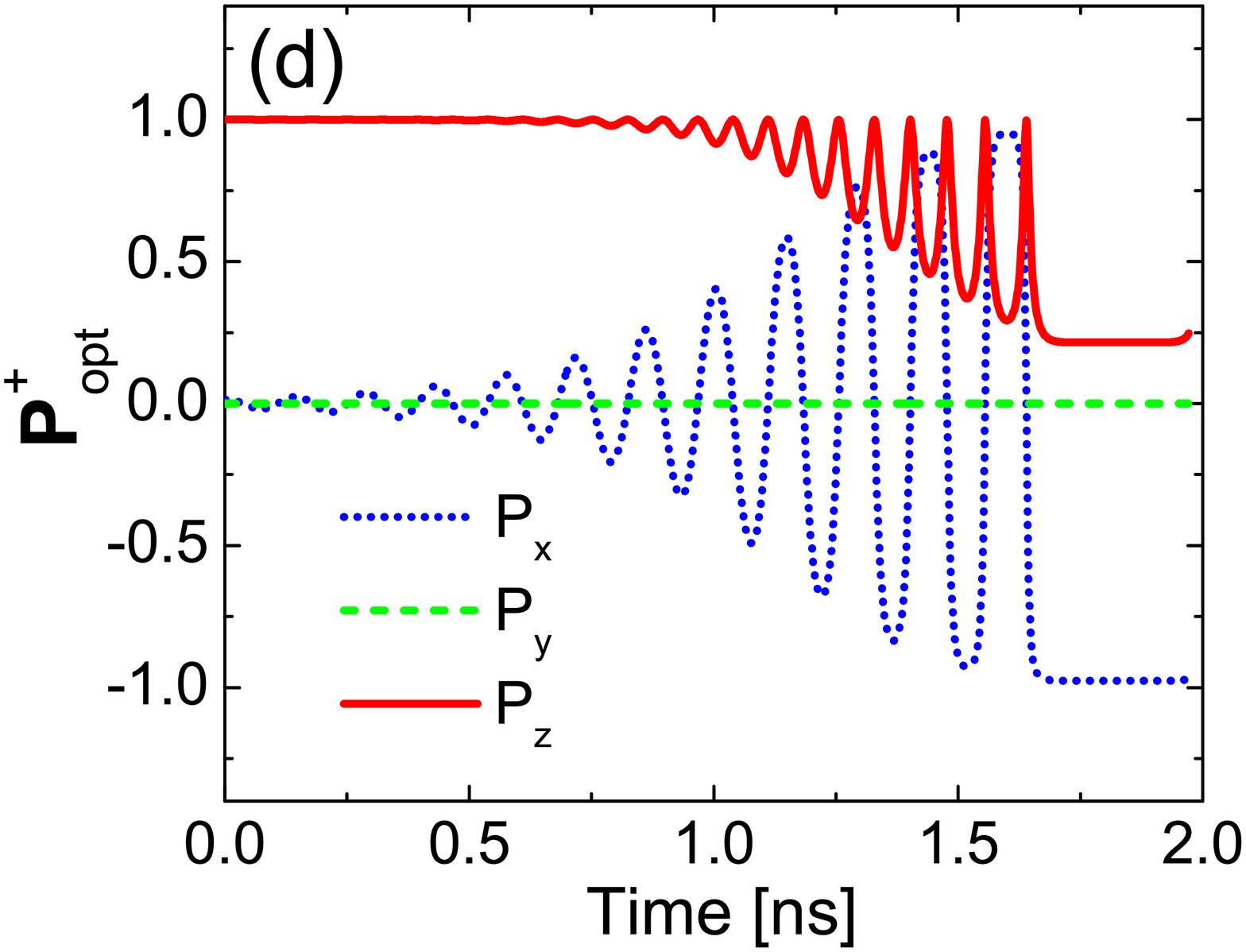}
\caption{(Color online) (a) optimal path $\{\mathbf{m}_{opt}^{+}(t)\}$ for the magnet to switch from $(0,0,1)$ to $(0,1,0)$; (b) a sample magnetization switching trajectory simulated by the stochastic LLG equation; (c) the amplitude of the optimal spin current pulse $I_{opt}^{s,+}$ corresponding to $\{\mathbf{m}_{opt}^{+}(t)\}$; (d) the spin polarization vector of the optimal spin current pulse $I_{opt}^{s,+}$ to enhance the happening probability of $\{\mathbf{m}_{opt}^{+}(t)\}$. }\label{optmP}
\end{figure}

The energy profile $E(\mathbf{m})$ has two saddle points, i.e. $(0,1,0)$ and $(0,-1,0)$, thus there exist two corresponding optimal paths for the magnet to switch from the initial stable point $(0,0,1)$ to the other $(0,0,-1)$. Fig.~\ref{optmP}(a) shows the optimal path $\{\mathbf{m}_{opt}^{+}(t)\}$ from $(0,0,1)$ to $(0,1,0)$, which is obained from Eq.~(\ref{OptPath}). As one can see, there are some oscillations around the stable point before the magnet arrives at the saddle point, and this process takes about 2~ns. For comparision, Fig.~\ref{optmP}(b) gives one random switching trajectory of the magnet simulated by the stochastic LLG equation (\ref{LLG}) with no spin current. It is found that the main feature of the real switching trajectories are indeed qualitatively captured by the optimal path, although there are some quanlitative differences because the noise amplitude is not small enough in our simulations. We can also see that the magnet randomly moves around the stable point for most of the time (about 1249~ns in this sample trajectory), and the switching event happens rather rarely. 

The optimal spin current pulse $I^{s,+}_{opt}(t)$ to enhance the happening probability of the optimal switching path $\{\mathbf{m}_{opt}^{+}(t)\}$ are given in Fig.~\ref{optmP}(c) and (d). The absolute value of $a_{J}^{opt} (t)$ is determined by the constraint condition, and its shape shows that the best control strategy is to allocate the spin current when the magnet has deviated from the stable point significantly. The optimal spin polarization direction $\mathbf{P}_{opt}^{+}$ is set to be in parallel with the effective magnetic field $\mathbf{H}_{eff}$. In contrast, the optimal spin current pulse to suppress the optimal path $\{\mathbf{m}_{opt}^{+}(t)\}$ should have an opposite spin polarization direction. For the energy profile (\ref{Eden}) considered here, $\mathbf{H}_{eff}$ has no $y$-component, and its $x$-component can be quite large even for a small $\mathbf{m}_{x}^{opt}$ due to the large demagnetization field. 

As a test, we calculated the mean switching time (MST) of the magnet under different control strategies,  where $10^{5}$ random trajectories are generated for each case. First, the MST without spin current is about $567$~ns. Then a constant spin current with $a_{J}=0.05a_{c}$ will increase the MST to be about $913$~ns if $\mathbf{P}=(0,0,-1)$, and decrease the MST to be about $353$~ns if $\mathbf{P}=(0,0,1)$. While for the optimal control, the corresponding spin current pulse $I^{s}_{opt}$ is applied only if the magnet randomly arrives at the critical position $\mathbf{m}$ that $\mathbf{m}_{z}=m_{c}$. We have chosen $m_{c}=0.89, 0.81, 0.65$ seperately, and the initial point of the corresponding spin current pulses are denoted as $1, 2, 3$ in Fig.~\ref{optmP}(c). The amplitudes $a_{J}^{opt}$ are given such that each pulse will contain the same energy as a $2$-ns constant pulse with $a_{J}=0.05a_{c}$. Thus, the number of pulses used directly reflects the energy consumption during the control process. 

\begin{table}[ht]
\caption{Mean switching times $\tau$ (unit : ns) and number of spin current pulses $n$ when the switching probability is either suppressed (s) or enhanced (e). $m_{c}$ : the critical values to activate the control spin current pulse. Subscript $a$ : the optimal spin current pulses $I_{opt}^{s}$ is applied; subscript $b$ : the spin polarization vector is fixed as $(0,0,-1)$ or (0,0,1) for suppressing or enhancing the switching respectively.} 
\centering 
\begin{tabular}{c |c  c|c  c} 
\hline
$\quad m_{c}\quad$ &\quad$\tau_{a}^{s}/n_{a}^{s}$\quad&\quad$\tau_{b}^{s}/n_{b}^{s}$\quad&\quad$\tau_{a}^{e}/n_{a}^{e}$\quad& $\tau_{b}^{e}/n_{b}^{e}$ \\
\hline 
0.89 & 517/430 & 1067/892 & 28/26 & 292/249  \\
0.81 & 739/263 & 1073/390 & 32/13 & 297/110 \\
0.65 & 515/32 & 903/57 & 67/4 & 347/22  \\ [1ex]
\hline
\end{tabular}
\label{taun} 
\end{table}

The resulting MSTs $\tau$ and mean pulse numbers $n$ are shown in Table~\ref{taun}, for both the cases to suppress and enhance the switching probability. Unfortunately, we found that the optimal pulses didn't increase the MSTs $\tau_{a}^{s}$ when $m_{c}=0.89$ and $0.65$. For $m_{c}=0.81$, the situation is better since the MST $\tau_{a}^{s}$ has been increased to $739$~ns with $263$ pulses, but it still has no obvious advantage compared with the constant control current case. The reason of the failure is that the noise amplitude is not small enough and the real switching paths ${\mathbf{m}(t)}$ can deviate from the optimal path significantly so that the control pulse will in fact increase their happening probabilities. To eliminate this effect, we have set the spin-polarization direction as $\mathbf{P}=(0,0,-1)$, and the results indeed become much better, as shown by $\tau_{b}^{s}$ and $n_{b}^{s}$ in Table~\ref{taun}.  Especially when $m_{c}=0.65$, the MST $\tau_{b}^{s}$ is $903$~ns which is close to the constant spin current case, but the number of pulses $n_{b}^{s}$ has been greatly decreased to $57$. This shows that the optimal control theory is helpful to find the way to save energy even when the noise amplitude is not small enough. 

The advantage of the optimal control strategy is fully verified for the case to increase the switching probability. In Table~\ref{taun}, we found that the MST $\tau_{a}^{e}$ is decreased to $28$~ns with only $26$ pulses for $m_{c}=0.89$. If $m_{c}=0.65$, the MST $\tau_{a}^{e}=66$~ns is larger, but merely $4$ pulses in average is used here. For comprison, we also list the results $\tau_{b}^{e}$ and $n_{b}^{e}$ if $\mathbf{P}$ is set as $(0,0,1)$ for the spin current pulses, which are only a little better than the constant spin current case. Obviously, the optimal control strategy successfuly increased the switching probability with greatly reduced energy consumption.   

In conclusion, we have discussed how to use the spin current to optimally control the stochastic magnetization dynamics based on the large deviation principle, and shown the advantage of the proposed control strategy by applying it to the thermal-driven magnetization switching problem. The happening probabilities of the rare events, which are the dominant part of the stochastic magnetization dynamics, are controlled by changing the corresponding quasi-potential with optimized spin current pulse. Further generalization of the idea to the cases including the multi-domain magnetization configuration, quantum fluctuations etc.  are necessary for more practical applications .

Y. W. thanks Dr. G. D. Chaves-O'Flynn for helpful discussions. This work is supported by the Hong Kong University Grant Council (AoE/P-04/08).


\begin{thebibliography}{99}
\bibitem{STT1} J. C. Slonczewski, J. Magn. Magn. Mater. \textbf{159}, L1 (1996).
\bibitem{STT2} L. Berger, Phys. Rev. B \textbf{54}, 9353 (1996).
\bibitem{STTrev1} D. C. Ralph and M. D. Stiles, J. Magn. Magn. Mater. \textbf{320}, 1190 (2008).
\bibitem{STTrev2} A. Brataas, A.D. Kent, and H. Ohno, Nature Materials \textbf{11}, 372 (2012). 
\bibitem{MagRev} J. W. Lau and J. M. Shaw, J. Phys. D: Appl. Phys. \textbf{44}, 303001 (2011).
\bibitem{Noise1} D. Weller and A. Moser, IEEE Trans. Magn. \textbf{35}, 4423 (1999).
\bibitem{Noise2} S. Urazhdin, N. O. Birge, W. P. Pratt, Jr., and J. Bass, Phys. Rev. Lett. \textbf{91}, 146803 (2003).
\bibitem{Noise3} R. H. Koch, J. A. Katine, and J. Z. Sun, Phys. Rev. Lett. \textbf{92}, 088302 (2004).
\bibitem{Noise4} I. N. Krivorotov, N. C. Emley, A.G. F. Garcia, J.C. Sankey, S. I. Kiselev, D. C. Ralph, and R. A. Buhrman, Phys. Rev. Lett. \textbf{93}, 166603 (2004).
\bibitem{Noise5} X. Cheng, C. T. Boone, J. Zhu, and I. N. Krivorotov, Phys. Rev. Lett. \textbf{105}, 047202 (2010). 
\bibitem{Exp} V. E. Demidov, S. Urazhdin, E. R. J. Edwards, M. D. Stiles, R. D. McMichael, and S. O. Demokritov, Phys. Rev. Lett. \textbf{107}, 107204 (2011).
\bibitem{LiZhang} Z. Li and S. Zhang, Phys. Rev. B \textbf{69}, 134416 (2004).
\bibitem{WangSham1} Y. Wang and L. J. Sham, Phys. Rev. B \textbf{85}, 092403 (2012).
\bibitem{Covington} M. Covington, U. S. Patent No. 7,042,685 (9 May 2006).
\bibitem{Bando} S.  Bandopadhyay, A. Brataas, and G.E.W. Bauer, Appl. Phys. Lett. \textbf{98}, 083110 (2011).
\bibitem{FW} M. Freidlin and A. Wentzell, \emph{Random Perturbations of Dynamical Systems} (Springer-Verlag, New York, 1998).
\bibitem{LDPrev} H. Touchette, Phys. Rep. \textbf{478}, 1 (2009).
\bibitem{Kohn} R. V. Kohn, M. G. Reznikoff, and E. Vanden-Eijnden, J. Nonlinear Sci. \textbf{15}, 223 (2005).
\bibitem{Kent} G. D. Chaves-O'Flynn, D. L. Stein, A. D. Kent, and E. Vanden-Eijnden, J. Appl. Phys. \textbf{109}, 07C918 (2011).
\bibitem{OpCon} V. N. Smelyanskiy and M. I. Dykman, Phys. Rev. E. \textbf{55}, 2516 (1997).
\bibitem{STT3} S. Zhang and Z. Li, Phys. Rev. Lett. \textbf{93}, 127204 (2004).

\end{thebibliography}
\end{document}